\documentstyle[aps,prb,twocolumn,graphics]{revtex}

\newcommand{\dfn}{\stackrel{\rm def}{=}}

\begin{document}

\title{Metal--insulator transitions in anisotropic 2d systems}
\author{Marc R{\"{u}}hl{\"{a}}nder and C. M. Soukoulis\\
Ames Laboratory and Department of Physics and Astronomy\\
Iowa State University, Ames, Iowa 50012}
\maketitle

\begin{abstract}
Several phenomena related to the critical behaviour of non--interacting
electrons in a disordered 2d tight--binding system with a magnetic field
are studied. Localization lengths, critical exponents
and density of states are computed using transfer matrix techniques.
Scaling functions of isotropic systems are recovered once the dimension
of the system in each direction is chosen proportional to the localization
length. It is also found that the critical point is independent of the
propagation direction, and that the critical exponents for the localization
length for both propagating directions are equal to that of the isotropic
system, $\nu \approx 7/3$. We also calculate the critical value $\Lambda_c$
of the scaling function for both the isotropic and the anisotropic system.
%It is found that $\Lambda_c^{\mathrm iso} = \sqrt{\Lambda_c^x \cdot \Lambda_c^y}$.
It is found that $\Lambda_c^{\rm iso} = \sqrt{\Lambda_c^x \cdot \Lambda_c^y}$.
Detailed numerical studies of the density of states $n(E)$ for the isotropic
system reveals that for an appreciable amount of disorder the critical
energy is off the band center.
\end{abstract}

\section{Introduction}

The problem of Anderson localization \cite{Krm93} in anisotropic sysytems
has attracted considerable attention \cite{LiQ97,Mil97,Zam96,Dup97} recently.
It is generally accepted \cite{LiQ97} that anisotropy does not change the
universality class and that the isotropic results are recovered once a
proper scaling of the anisotropic results is performed. If the dimension
of the system size is chosen to be directly proportional to the localization
length, the system should be effectively isotropic. The difficulty in
implementing such a procedure lies in the fact that the localization lengths
are usually not known a priori. It was found through detailed numerical
calculations \cite{LiQ97} that this scaling indeed works. It was also shown
\cite{Wan97} that the probability distributions of the conductance in the
two directions are exactly equal to each other, provided that the ratio of
the sides of the rectangle is proportional to the ratio of the localization
lengths in the two directions. These scaling results were obtained for an
anisotropic system where all the states were localized.

It is well known \cite{Krm93} that non--interacting electrons are localized
in 2d disordered systems. There are, however, some exceptions to this rule.
These include electrons having strong spin--orbit coupling \cite{Hik80},
integer quantum Hall systems \cite{Huc95}, and tight binding models
with random magnetic fields \cite{Fur99}. The best known example is
the integer quantum Hall plateau transition occuring in a 2d
non--interacting system in a strong magnetic field. Extended states
do not exist as a result of Anderson localization except at a singular
energy near the center of each of the Landau subbands \cite{Huc95,Wan98b}.
The localization length diverges at these critical energies $E_c$ as
$\xi \propto | E - E_c |^{-\nu}$.

Another important point is the universality of the conductance at the
critical point of the Anderson transition for the anisotropic 
system\cite{Zam96}. From the generalized
scaling functions, it has been established that the geometric mean of the
the critical value $\Lambda_c$ of the scaling function
$\Lambda \dfn (\lambda_M/M)$ (as a function of $\xi/M$) is a 
constant independent of the strength of the anisotropy. 
(Here $\lambda_M$ denotes the finite size localization length of a 
quasi--one--dimensional strip of finite width $M$.) Numerical
calculations in both two \cite{LiQ97} and three \cite{Zam96} dimensional
disordered anisotropic systems support this claim. However, the same is
not true for the conductance. Numerical calculations \cite{Zam96} in three
dimensional anisotropic systems do not support a universal value of the
conductance for the geometric mean. This might be due to too small sizes
used in the 3d system or to a lack of universality of the conductance at
the critical point.

In this paper we investigate the scaling properties of the finite size
localization length $\lambda_M$ and the critical value $\Lambda_c$
of the scaling function
in a two--dimensional system described by a tight binding model in the
presence of a magnetic field. 
Both the isotropic case, as well as the anisotropic case will be examined.
This is perhaps the simplest system that
exhibits the correct behaviour of the metal to insulator transition.
To our knowledge, no such calculations have been previously reported for
the anisotropic tight binding model with a constant magnetic field.
Some of the questions we try to answer are: how does the anisotropy
effect the critical behaviour, especially, will there be one or two
critical exponents for the localization lengths; how do the anisotropic
quantities relate to the corresponding isotropic ones, especially, can
we expect the geometric mean of the two anisotropic values to equal 
the isotropic value; and what are the values for the scaling functions
at the critical point? In section \ref{meth} we describe the model and
the numerical methods we used, in section \ref{results} we present
and discuss our numerical results and in section \ref{summ} we
summarize the conclusions of this work.

\section{Model and Methods}
\label{meth}

In the tight binding model one has the Hamiltonian

\begin{equation}
%\mathcal{H} = \sum_i \left| i \right\rangle \varepsilon_i \left\langle i 
{\cal H} = \sum_i \left| i \right\rangle \varepsilon_i \left\langle i 
\right| + {\sum_{i,j}}' \left| i \right\rangle V_{ij} \left\langle j \right|
\end{equation}

\noindent
where the summations run over lattice sites $i$ and $j$. We consider
only nearest neighbour interaction in the hopping integrals $V_{ij}$.
The effects of an external magnetic field, characterised by a vector
potential $\mathbf{A}$ ($\nabla \times \mathbf{A} = \mathbf{B}$), enter
the model via phases of the hopping integrals with

\begin{equation}
V_{ij} = t_{ij} e^{-2\pi i \frac{e}{h} \int_{\mathbf{r}_i}^{\mathbf{r}_j}
\mathbf{A}(\mathbf{r}) \mathrm{d}\mathbf{r}}
\end{equation}

\noindent
the integral connecting lattice sites $i$ and $j$ by a straight line.
In two dimensions with a magnetic induction $\mathbf{B}$ perpendicular
to the plane of the system, one can choose the gauge for the vector
potential in such a manner that the phases vanish in one direction
within the plane and are integer multiples of some number $2\pi\alpha$
in the other direction,
such that the value of $\alpha$ completely characterises the influences
of the magnetic field on the system. In particular, the denominator of
a rational $\alpha$ equals the number of bands in the density of states
of the system without disorder. Introducing anisotropy into the system 
by choosing different amplitudes $t_{ij}$ in the two directions within 
the plane will effect only the position of these bands, not their
number. We bring disorder into the system by independently choosing all
the site energies $\varepsilon_i$ from a rectangular distribution of
width $W$ centered at $0$; thus $W$ is a measure of disorder strength.
Both $W$ and $E$ are measured in units of the largest hopping matrix
element $t$, which is taken to be unity.

As our main method we use the transfer matrix method \cite{Krm93}, where
a matrix
${\mathrm T}_M$ connects the amplitudes of a state at both ends of a 
quasi--one--dimensional strip of width $M$ and length $N \gg M$.
Due to the anisotropy, we have to do this in the two spatial directions
seperately. Therefore we get two sets of parameters $\lambda_{M,x}$ and
$\lambda_{M,y}$ which lead to two seperate localization lengths $\xi_x$
and $\xi_y$ in the $x$-- and $y$--direction respectively. 
Scaling of the data
is used to improve on the values of $\xi_x$ and $\xi_y$, which are then
analysed to find the critical energies, where the localization lengths
diverge as well as the critical exponents of these divergences.

We obtain the density of states by using a Lanczos procedure \cite{Cul85} to
diagonalize the Hamiltonian on squares of (linear) size $L$. The
energy level separation distribution function $P(s)$ should deviate
markedly from a Poisson distribution for the local density of states
around the critical energy, approaching the Wigner distribution for
the unitary ensemble \cite{Boh84,Meh91}. We also use the density of states to show that
for sufficiently strong disorder the critical energy does not necessarily
coincide with the band center.

\section{Results}
\label{results}

\begin{figure*}[t]
\resizebox{3.0in}{3.0in}{ \includegraphics{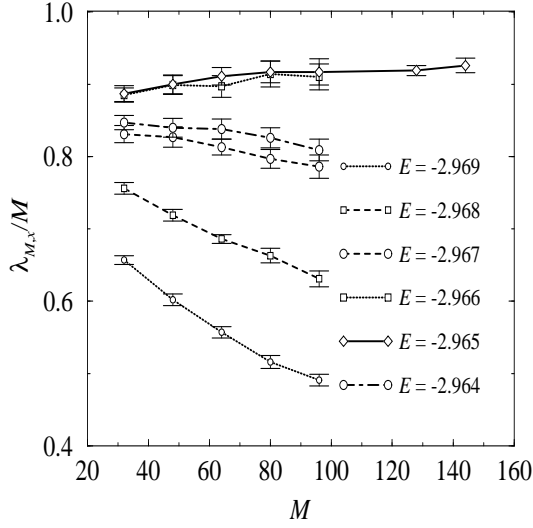} }
\resizebox{3.0in}{3.0in}{ \includegraphics{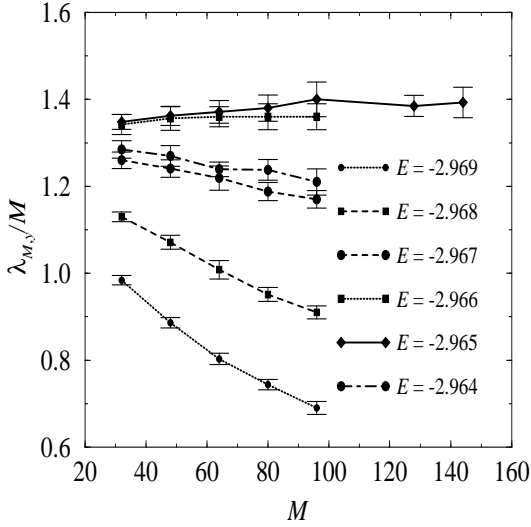} }
\caption{\label{lofM} The critical values of the scaling functionns can be
obtained from the large--$M$ limit of $\lambda_M/M$ at the critical energy.
From the almost symmetrical behaviour of the values for the non--critical
energies at either side of the critical one, we assume that the value for
$E_c$ is between $-2.966$ and $-2.965$, but closer to the second one. Top:
difficult hopping direction; bottom: easy hopping direction.}
\end{figure*}

To obtain the critical energy $E_c$ for the anisotropic tight binding
model, first we calculate $\lambda_{M,x}$ and $\lambda_{M,y}$ for different
strip widths $M$ and energies $E$ above and below the critical energy
$E_c$. As the exact position of $E_c$ varies with the disorder strength
$W$, the hopping integral $t_x < 1$ in the difficult hopping direction
as well as the magnetic field parameter $\alpha$, we restrict our 
investigation to one set of these parameters $W = 0.1$, $t_x = 0.8$ and
$\alpha = \frac{1}{8}$. The data for the more localised states show that 
$M/\lambda_M$ versus $M$ is a straight line. The inverse slope of each of these
lines gives a first estimate for the localization lengths $\xi_x$ or
$\xi_y$ respectively, thus the smaller the slope, the more extended are the
corresponding eigenstates of the system. For energies closer to $E_c$
the lines would be essentially horizontal. In order to
accurately obtain $E_c$, we have systematically calculated $\lambda_{M,x}$
and $\lambda_{M,y}$ for large $M$.
The results are shown in Fig.\ \ref{lofM}, where we plot $\lambda_{M,x}$ and 
$\lambda_{M,y}$ for the anisotropic case 
for energies very close to $E_c$. 
From Fig.\ \ref{lofM} we can confirm the existance
of an extended state. Notice that $\lambda_M/M$ decreases as a function of
$M$, which signifies localized states. For localized states, $\lambda_M$
eventually reaches its large--$M$ limit, which is a constant, and therefore
$\lambda_M/M$ decreases as $M$ increases. However, as can be seen from
Fig.\ \ref{lofM}, at the critical energy $E_c$, $\lambda_M/M$ saturates
to a constant due to the absence of length scales. For the case studied
($W = 0.1$, $t_y = 1.0$, $t_x = 0.8$ and $\alpha = 1/8$) we find that
the critical energy $E_c$ is between $-2.966$ and $-2.965$, but closer
to the second value. From Fig.\ \ref{lofM} we can also obtain the critical
values $\Lambda_c$ of $\lambda_M/M$ for both directions of propagation.
We find $\Lambda_c^x = 0.92 \pm 0.01$ and $\Lambda_c^y = 1.39 \pm 0.01$
with a geometric mean of $1.13 \pm 0.01$.

\begin{figure}[h]
\resizebox{3.0in}{3.0in}{ \includegraphics{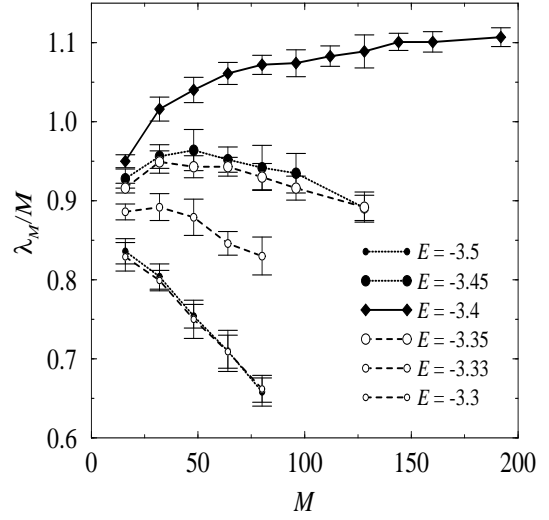} }
\caption{\label{lofM-iso} The critical value of the isotropic scaling
function. From the large--$M$ data we estimate it to be $1.10 \pm 0.03$.}
\end{figure}

To confirm that the geometric mean $\Lambda_c$ of the two anisotropic
values is related to the value for the isotropic case, we have
calculated systematically $\lambda_M/M$ versus $M$ for the isotropic
system ($W = 4.0$, $t_x = t_y = 1.0$ and $\alpha = 1/8$) for very
large values of $M$. These results are shown in Fig.\ \ref{lofM-iso}.
From Fig.\ \ref{lofM-iso} we obtain that indeed $E_c = -3.40$ in this case,
in agreement with previous results \cite{Wan98b,Pot98a} that used
different techniques to get $E_c$. In addition, Fig.\ \ref{lofM-iso}
shows clearly that at the critical point of the isotropic system
$\Lambda_c^{\mathrm iso} = 1.10 \pm 0.03$, which is approximately equal to the
geometric mean of the two anisotropic values $\Lambda_c^x$ and $\Lambda_c^y$.

The critical value of $\lambda_M/M$ is related to the exponent $\alpha_0$
that can be obtained from the multifractal analysis\cite{Huc95} of the 
eigenfunctions at the critical energy by $\Lambda_c^{-1} = \pi (\alpha_0 - d)$ 
where $d$ is the Euclidian dimension of the system. Huckestein \cite{Huc94}
calculated $\Lambda_c = 1.14 \pm 0.02$ for a real space model, while
Lee et al.\ \cite{Lee93} determine $\Lambda_c = {\mathrm ln}^{-1} (1 +
\sqrt{2}) \approx 1.13$ for a network model, both of which are close
to the value $\Lambda_c^{\mathrm iso} = 1.10 \pm 0.03$ obtained
for the isotropic case of the 2d tight binding model with a constant
magnetic field.

\begin{figure}[h]
\resizebox{3.0in}{3.0in}{ \includegraphics{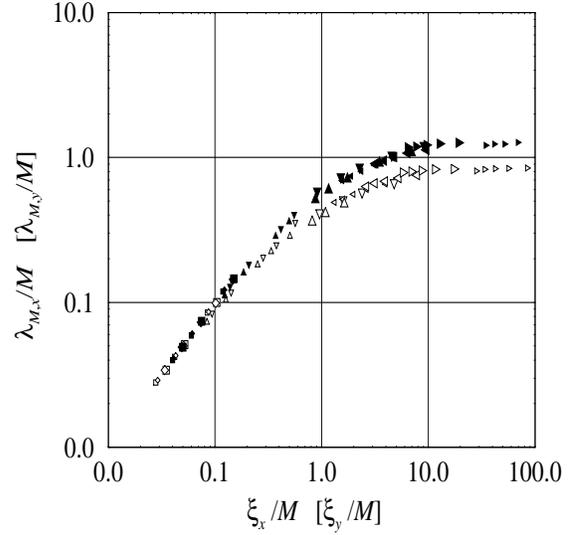} }
\caption{\label{lofx} The scaling functions for the difficult (open symbols)
and easy (filled symbols) hopping directions $\lambda_M/M$ as a function of
$\xi/M$. The localization lengths have
been adjusted to better fit the data to a smooth curve. Energies are
$-3.0$, $-2.99$, $-2.98$, $-2.97$, $-2.969$, $-2.968$, $-2.967$, $-2.964$,
$-2.96$, $-2.95$, $-2.94$ and $-2.93$.}
\end{figure}

\begin{figure}[h]
\resizebox{3.0in}{3.0in}{ \includegraphics{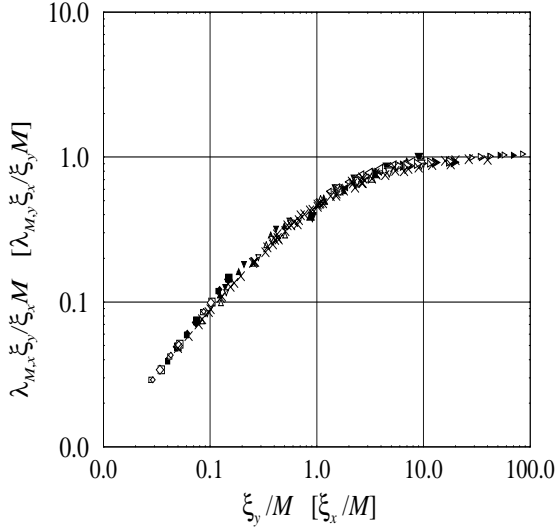} }
\caption{\label{lofx2} Plotting the rescaled scaling functions (cf. text)
for the difficult (open symbols) and easy (filled symbols) hopping directions
together with the scaling function for an isotropic system (crosses).}
\end{figure}

The next step is to use the values for the localisation lengths obtained
in this manner to plot $\lambda_{M,x}/M$ as a function of $\xi_x/M$ and
$\lambda_{M,y}/M$ as a function of $\xi_y/M$. After combining the data for
all energies into one graph, one usually has to adjust the values for the
localization lengths slightly to make the data fall on a smooth curve.
Fig.\ \ref{lofx} shows that these two functions are independent of the
value of $E$, as expected for one--parameter scaling. However, the two
scaling functions differ in their large--$\xi$ limit: the value is higher
for the easy--hopping direction.
To compensate for this anisotropy effect we use the following
straightforward idea \cite{LiQ97}: $\lambda_{M,x}$ ($\lambda_{M,y}$) is a length
in the $x$-- ($y$--) direction along the length of the strip, so
the appropriate scale should be $\xi_x$ ($\xi_y$). However, $M$ is
a length measuring the width of the strip and therefore has to be scaled 
with the other localization length. Thus we plot $\frac{\lambda_{M,x}}{\xi_x}
\cdot \frac{\xi_y}{M}$ vs $\frac{\xi_y}{M}$ and $\frac{\lambda_{M,y}}{\xi_y}
\cdot \frac{\xi_x}{M}$ vs $\frac{\xi_x}{M}$ in Fig.\ \ref{lofx2}.
Not only do we obtain the same scaling function for both, but it is
also the same as the isotropic one which we included for reference.
The isotropic case was for $W = 4$ and $\alpha = 1/8$. Of course under the
assumption of one parameter scaling the form of the (isotropic) scaling 
function should not depend on the values of $W$ and $\alpha$ directly
(as long as neither vanishes completely) but only parametrically via
the loclaization length $\xi(E,W,\alpha)$.
Thus, the product of the two rescaled anisotropic functions equals the
square of the isotropic scaling function.
Immediately it is seen from this that the isotropic scaling function
equals the geometric mean of the two anisotropic scaling functions

\begin{equation}
\label{geomean}
%\left( \frac{\lambda_M}{M} \right)_\mathrm{iso} = 
\left( \frac{\lambda_M}{M} \right)_{\rm iso} = 
\sqrt{ \frac{\lambda_{M,x}}{M} \cdot \frac{\lambda_{M,y}}{M} }
\end{equation}

\noindent
as the rescaling factors $\xi_x$ and $\xi_y$ cancel each other.
As we have shown before in Fig.\ \ref{lofM} and Fig.\ \ref{lofM-iso},
indeed Equ.\ \ref{geomean} is obeyed.

\begin{figure}[h]
\resizebox{3.0in}{3.0in}{ \includegraphics{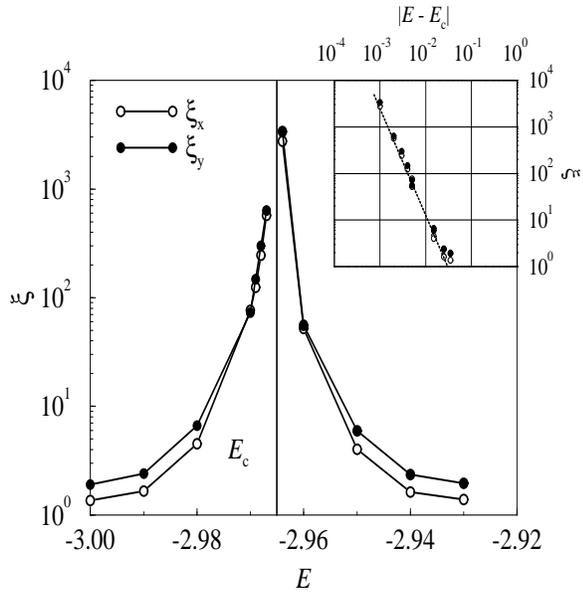} }
\caption{\label{xofE} The divergence of the localization lengths at
the critical Energy; open symbols: difficult hopping direction,
filled symbols: easy hopping direction. The values are taken after the 
adjustments made to obtain Fig.\ \ref{lofx2}. Inset:
\label{xofE-log} To extract the critical exponent of the localization
lengths we plot $\xi$ vs $\left| E - E_c \right|$ in a log--log plot.
Both exponents are found to be $2.3 \pm 0.1$, roughly equal to the theoretical
value for the isotropic system.}
\end{figure}

The procedure of fitting the data to a smooth scaling function
provides us with more accurate estimates of the localization lengths
which we can now use to determine the critical behaviour of $\xi$.
In Fig.\ \ref{xofE} we plot the localization lengths as a function of
energy. One can clearly see that the states are less localized in the
easy hopping direction, as was to be expected. Fig.\ \ref{xofE} also
allows us to estimate $E_c$, the energy where the localization length 
diverges. We expect this critical energy to be independent of the
strip orientation, as a higher--dimensional system would undergo a phase 
transition at this point, and our data give a strong indication that $E_c$
is indeed the same for both directions. We estimate $E_c \approx -2.965
\pm 0.001$. This is consistent with the results obtained in Fig.\ \ref{lofM}.

The divergence of the localization length near the critical energy
is expected to follow a power law

\begin{equation}
\xi(E) = \xi_0 \left| E - E_c \right|^{-\nu}
\end{equation}

\noindent
with some critical exponent $\nu$. To test this hypothesis, we plot
the logarithm of $\xi$ vs the logarithm of $\left| E - E_c \right|$.
The result is shown in the inset of Fig.\ \ref{xofE-log}. That our data follows
a straight line rather reasonably reconfirms our estimate for $E_c$,
as the plot obviously is quite sensitive to the choice for that value.
Furthermore, both sets of data can be fitted by the same straight line,
giving the same critical exponent $\nu \approx 2.3 \pm 0.1$. Once again,
this is the same as the isotropic value and very close to the
theoretically predicted value \cite{Huc95} of $7/3$ for the isotropic system.

The distribution of energy level separations in a given
energy interval depends on the typical extension of the eigenstates
of the system with eigenvalues in that energy region. 
Spatial overlap of eigenfunctions close in energy helps delocalizing
the particle. In a finite system, more of the strongly localized
eigenfunctions can be accomodated without significant overlap.
The more extended the eigenfunctions become, it gets more and more 
difficult to fit several into the finite space and they must be seperated
in energy. This leads to the phenomenon of level repulsion, known
from chaos theory. The corresponding distribution of level separations
$s_i = E_i - E_{i-1}$ goes to zero for small $s$. In contrast, the
distribution for a range of localized eigenstates has a maximum at
vanishing level separation. More specifically, Random Matrix Theory 
predicts\cite{Meh91} a Poisson distribution for the localized case, and a Wigner
distribution for the extended case. We have calculated the distribution
of energy level separations $p(s)$ for the anisotropic system studied
in Fig.\ \ref{lofM}. We find that for an energy range close to $E_c = -2.965$,
$p(s)$ is Wigner--like, whereas for the other energy ranges it is Poisson--like.

Level statistics for the isotropic 
system have been extensively studied by Potempa et al. \cite{Pot98a,Pot98b}
and Batsch et al. \cite{Bat97,Bat98}, proving the validity of the
approach in distinguishing localized from extended states. 
In addition, the level number variance $\Sigma^2(\langle N \rangle)
= \chi \cdot \langle N \rangle$ has been numerically obtained for the isotropic
system \cite{Kle97}, using the Chalker-Coddington network model \cite{Cha88},
and compared with analytical theories \cite{Cha96} which
give for the spectral compressibilty $\chi = (d - D(2))/2d$, where $D(2)$
is the multifractal exponent of the wavefunction at the critical point
\cite{Jan94}. Klesse and Metzler obtain $\chi = 0.124 \pm 0.006$ \cite{Kle97}.
Numerically obtained values for $D(2)$ include $1.43 \pm 0.03$ for a
continuum model \cite{Poo91}, $1.56$ for a network model \cite{Huc98},
and $1.62 \pm 0.02$ and $1.71$ for a tight binding model \cite{Huc92a,Huc94a}.
Due to the limited size of our systems we were not able
to produce results for $\chi$ for our anisotropic model.
The number of energy eigenvalues sufficiently close to the critical
point is not large enough to give good statistics for the number variance
$\Sigma^2(\langle N \rangle)$. This point has to be adressed in the future.

\begin{figure}[h]
\resizebox{3.0in}{3.0in}{ \includegraphics{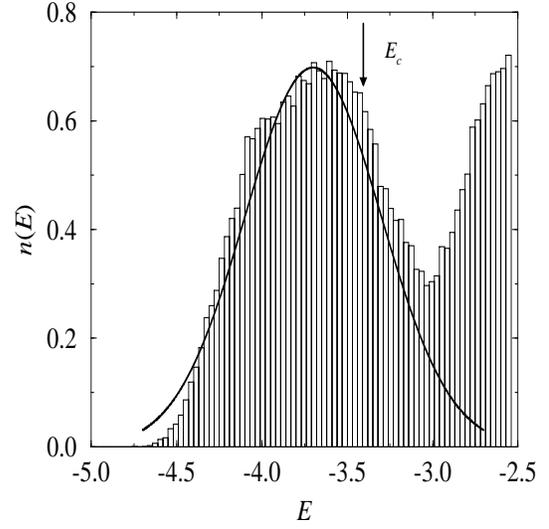} }
\resizebox{3.0in}{3.0in}{ \includegraphics{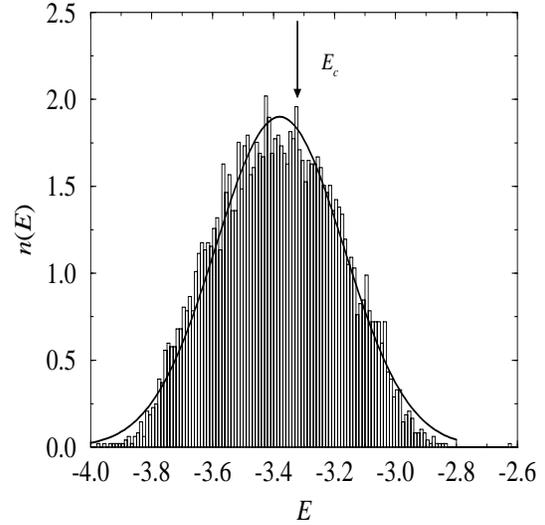} }
\caption{\label{dos} The density of states for the lowest subband for a
disorder strength of $W = 4.0$ (top) and $W = 2.0$ (bottom), indicating that
the critical energy is off the band center. A fit to a gaussian distribution
suggests that the band center in the $W = 2.0$ case is at $E_0 \approx -3.38$,
whereas the critical energy is $E_c \approx -3.32$.}
\end{figure}

Finally, Fig.\ \ref{dos}
shows the positions of $E_c$ for isotropic systems at $W=2$ and $W=4$
to be different from the band center.
Although a Gaussian is not the correct form for the density of states
it is usually a reasonable fit. For the stronger disorder, $W=4$, the
best approximation is achieved with a gaussian centered at $E = -3.7$
with a standard deviation of $\sigma = 0.4$. Fig.\ \ref{lofM-iso} strongly
indicates $E_c = -3.4$ (arrow in top panel of Fig.\ \ref{dos}).
Similarly, for the lesser disorder, $W = 2$, a gaussian centered at
$E = -3.38$ with a standard deviation of $\sigma = 0.21$. A plot
similar to that for the more strongly disordered case shows that
$E_c = -3.32$. However, for $W \leq 1$ the critical energy $E_c$ lies
at the center of the Landau band.

\section{Conclusions}
\label{summ}

In summary, we have performed detailed numerical study of the scaling
properties of highly anisotropic systems in 2d, with a metal to insulator
transition. Scaling functions of the isotropic systems are recovered
once the dimension of the anisotropic system is chosen to be proportional
to the localization length. It is also found that the critical point is
independent of the propagation direction and that the critical exponents
for the localization length in both propagating directions are equal
to that of the isotropic system. The critical value $\Lambda_c$ of the
scaling function for both the isotropic and the anisotropic cases has been
%calculated. It is obtained that $\Lambda_c^{\mathrm iso} = \sqrt{\Lambda_c^x
calculated. It is obtained that $\Lambda_c^{\rm iso} = \sqrt{\Lambda_c^x
\cdot \Lambda_c^y} = 1.10 \pm 0.03$. Finally, density of states calculations
revealed that the critical energy lies away from the center of the Landau
band.

\section*{Acknowledgements}

Ames Laboratory is operated for the U.S.\ Department of Energy
by Iowa State University under Contract No.\ W--7405--Eng--82.
This work was supported by the Director for Energy Research, Office
of Basic Sciences.

%\bibliography{Bib/references}

\end{document}